**PRACTICE PAPER**

# 39 Hints to Facilitate the Use of Semantics for Data on Agriculture and Nutrition

Caterina Caracciolo[1], Sophie Aubin[2], Clement Jonquet[3], Emna Amdouni[3], Romain David[4,5], Leyla Garcia[6], Brandon Whitehead[7], Catherine Roussey[8], Armando Stellato[9] and Ferdinando Villa[10]

[1] Food and Agriculture Organization of the United Nations (FAO), HQ Rome, IT
[2] DipSO, INRAE, Angers, FR
[3] LIRMM, University of Montpellier and CNRS, FR
[4] MISTEA, University of Montpellier, INRAE, Institut Agro, Montpellier, FR
[5] ERINHA (European Research Infrastructure on Highly Pathogenic Agents) AISBL, Paris, FR
[6] ZB MED Information Centre for Life Sciences, Cologne, DE
[7] Manaaki Whenua – Landcare Research, Palmerston North, NZ
[8] TSCF, INRAE, Aubière, FR
[9] University of Rome Tor Vergata, Rome, IT
[10] Basque Centre for Climate Change (BC3); IKERBASQUE, Basque Foundation for Science, ES
Corresponding author: Caterina Caracciolo (caterina.caracciolo@fao.org)

In this paper, we report on the outputs and adoption of the Agrisemantics Working Group of the Research Data Alliance (RDA), consisting of a set of recommendations to facilitate the adoption of semantic technologies and methods for the purpose of data interoperability in the field of agriculture and nutrition. From 2016 to 2019, the group gathered researchers and practitioners at the crossing point between information technology and agricultural science, to study all aspects in the life cycle of semantic resources: conceptualization, edition, sharing, standardization, services, alignment, long term support. First, the working group realized a landscape study, a study of the uses of semantics in agrifood, then collected use cases for the exploitation of semantics resources – a generic term to encompass vocabularies, terminologies, thesauri, ontologies. The resulting requirements were synthesized into 39 "hints" for users and developers of semantic resources, and providers of semantic resource services. We believe adopting these recommendations will engage agrifood sciences in a necessary transition to leverage data production, sharing and reuse and the adoption of the FAIR data principles. The paper includes examples of adoption of those requirements, and a discussion of their contribution to the field of data science.

## 1. Introduction

Data are important in agriculture, including fields such as precision agriculture (Shannon et al., 2020), climate modeling (Crosson et al., 2011; Nelson et al., 2014) and policy making. There is a clear need for better ways to generate, access, exploit and reuse data (Platform for Big Data in Agriculture, 2018, 2017) across different information systems. Agricultural and food data, similar to other domains, should fully support and implement the Findable, Accessible, Interoperable, and Reusable (FAIR) principles (Wilkinson et al., 2016). Efforts such as the Rothamsted long term agriculture experiment (Johnston and Poulton, 2018; Perryman et al., 2018; Poulton et al., 2018), the CGIAR Big Data platform (CIAT, n.d.), the Global Open Data for Agriculture and Nutrition (GODAN) organization (Musker et al., 2018; Musker and Schaap, 2018), and the AgBioData consortium (Harper et al., 2018) illustrate the value of data in agriculture and food systems. Given the importance of data interoperability to researchers and practitioners working at the interface between data and information management and agriculture, we formed the Agrisemantics Working Group (WG) within the Research Data Alliance (RDA).[1]

The Agrisemantics WG, active between 2016 and 2019, is now in "maintenance phase". It was created as a community-driven initiative, formed for the purpose of understanding the specific features and needs of the community and identifying solutions that best fit. The group was headed by representatives of three of the major organizations in agriculture research, CABI, INRAE and FAO, respectively B. Whitehead, S. Aubin, C. Caracciolo; it registered 120 people, with about 30 active members over the years. Coordination was ensured through regular calls, face-to-face meetings during the RDA plenary meetings, and co-located events of the RDA Agricultural Data Interest Group[2] (IGAD). The group organized its work around three main activities: 1) investigating the current use of semantics in the area of agriculture; 2) surveying the community on bottlenecks for use (and re-use) of semantics in agriculture, along with any solutions under development in the area; and 3) producing a set of recommendations relative to the domain.

The Agrisemantics WG focused on solutions and recommendations based on semantics, i.e., the study of meaning from the knowledge engineering perspective. Along this text, we use "semantics" to refer to scientific methods and technologies enabling semantic representation and exploitation of data (Domingue et al., 2011; Hitzler et al., 2010) and by "Semantic Web technologies" we refer to the set of languages and formats enabling publication of linked data on the web. The meaning of data is commonly expressed by semantic resources (aka semantic structures, or semantic artifacts, or more generally knowledge organization systems (Zeng, 2008)), typically sets of terms and definitions organized in ways that reflect views of the world adopted when collecting the data, suitable to the intended applications of the data. Such structures vary from flat to hierarchical (taxonomies), to more complex structures supporting reasoning (ontologies). A semantic resource (SR from now on) may be used to provide the "qualitative data dimension" (the findable reference id representing the real entity measured in statistical data), defining what a numeric value actually means; values for metadata elements (the scope of a dataset or its provenance information), usually collected into metadata sets (or schemes); or to formalize data dictionaries.

We argue that semantics is key for data interoperability. Providing explicit and machine-readable[3] description of data makes it possible to programmatically integrate and reuse data. With this in mind, the Agrisemantics WG formulated two questions (Aubin et al., 2017b): Is there a specific semantics for agriculture? Is there a specific type of interoperability for agricultural data? Therefore, we focused on semantics for agricultural data, understood as "data produced or used in agriculture and food systems, including data on agricultural production, or agronomic data relative to lab and field experiments, environmental conditions, soils or climate, just to mention a few relevant areas of data productions" (Aubin et al., 2017b).

This paper is organized as follows: Section 2 summarizes the work done by the Agrisemantics WG reporting on the current status of semantics and the corresponding bottlenecks in the agriculture domain, and the final output of the group, the "39 Hints" (in full in Annex I). Section 3 discusses the current and future adoption of the Agrisemantics recommendations. Finally, Section 4 draws some conclusions from the group activities.

## 2. Activities of the Agrisemantics WG

Here we summarized the previous deliverables produced by the Agrisemantics WG as they act as foundational elements for the adoption cases reported on the next section. The main findings of the investigation on the use of semantics resources in general and then tailored to the agricultural data (Aubin et al., 2017b) are discussed in Section 2.1. The second activity (Section 2.2) corresponds to a survey regarding the needs while working with semantic resources (Aubin et al., 2018). The third activity (Section 2.3) summarizes the 39 hints to facilitate the use of semantics for data on agriculture and nutrition (RDA Agrisemantics WG, 2019).

---

[1] RDA's Agrisemantics WG's page is: https://www.rd-alliance.org/groups/agrisemantics-wg.html but the group has also consolidated outputs on another dedicated web site to foster further discussion: https://agrisemantics.org/.
[2] https://www.rd-alliance.org/groups/agriculture-data-interest-group-igad.html.
[3] Not only digital but also expressed in formats that can be automatically read and interpreted by a computer program without the need for manual human intervention.

### 2.1 Use of semantic resources

Five main common data-related tasks regularly involve semantics: search, information extraction, data models, data integration and automated reasoning (**Figure 1**). **Data search** involves the use of indexes associated with keywords, either free or from semantic resources. Semantic keywords make it easier to avoid mistakes due to misspelling and enable the use of synonyms and possibly multilingual approaches. **Information extraction**, commonly done via text-mining, facilitates identifying elements and relations in the text that can be associated with controlled vocabularies, i.e., named entity recognition. **Data models** act as a conceptualization of reality based on entities and their relations; a semantic data model relies on meanings, making FAIR easier to realize. **Data integration** refers to the harmonization and reuse of data created by others preserving the meaning when moving across systems. Ontologies play a major role for data integration as they can provide rules to transform and harmonize data in a meaningful manner. **Automated reasoning** makes it possible to find out non-explicit content, leading to the discovery of new facts and knowledge.

Based on the consideration that semantics is ubiquitous in tasks involving data, the group investigated the status of formal semantic resources for the agriculture domain. To this end, we used: a "Map of data standards for food and agriculture" (from now on "the Map") corresponding to a catalog of SRs and standards that later became the "Agrisemantics Map of Data Standards" (Pesce et al., 2018), and the AgroPortal vocabulary and ontology repository (Jonquet et al., 2018b, 2018a). To the best of our knowledge, the Map and AgroPortal are the only initiatives specifically addressing the agrifood domain. We noticed that while controlled vocabularies are commonly expressed following the expected W3C Recommendations, e.g., OWL, RDFS and SKOS, a few SRs use other standards such as OBO or specific tabular data formats (e.g., Crop Ontology TDv5). Glossaries and classification schemes often use XML or non-fully machine-readable formats. Regarding metadata, the situation is precarious as metadata elements useful to the user are usually missing (for example, no links to documentation web pages or license). Overall, nearly half of the resources used to describe agricultural data (e.g., controlled vocabularies or classification systems) are in PDF – so digital, but not fully machine-processable. The area of plant sciences seems to be the one for which the most SRs have been developed (Cooper et al., 2018; Pommier et al., 2019). The reader can also refer to (Harper et al., 2018) for an overview of use of ontologies in different agriculture databases and (Drury et al., 2019) for a review of semantic resources in agriculture and examples of their application. The area of plant sciences seems to be the one for which the most SRs have been developed (Cooper et al., 2018; Pommier et al., 2019). The reader can also refer to (Harper et al., 2018) for an overview of use of ontologies in different agriculture databases and (Drury et al., 2019) for a review of semantic resources in agriculture and examples of their application.

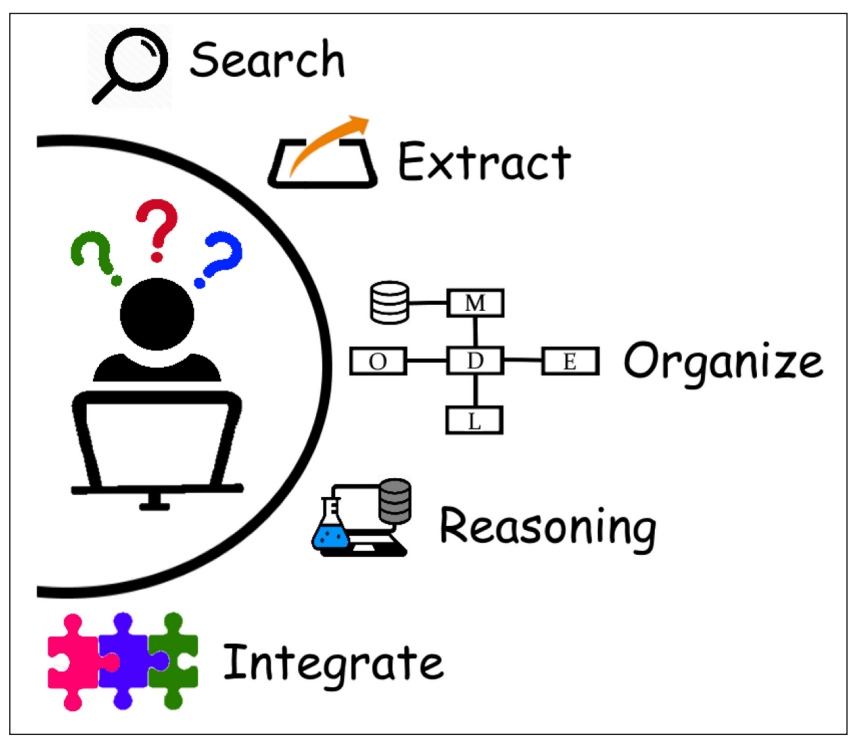

**Figure 1:** The five main tasks identified by the Agrisemantics WG as involving semantics.

## 2.2 Survey: "What are the Needs When Working with Semantic Resources?"

Given the fact that only few semantic resources are available in standard and machine-readable format, with expected consequences on the data interoperability and the activity of data researchers and practitioners, the group moved on to investigate real-life problems and bottlenecks that researchers and practitioners encounter when working with data and SRs, together with their wishes and/or proposed solutions. We set up an open-ended question survey and a template to gather information[4] (Aubin et al., 2017a; Baker et al., 2017; Caracciolo et al., 2019a). We asked for information on open problems, and ideas for solutions at different stages of development, including ongoing and future projects. We collected 20 use cases,[5] from institutions based in 10 distinct countries; 15 from Europe, 2 from North and 2 from South America, 1 from Asia. They were primarily from research organizations (15), with a handful from international (3), professional (1) and governmental (1) organizations.

We found that different roles and backgrounds are involved in different tasks dealing with SRs, showing that the process of producing SRs is highly collaborative and requires various competencies. For instance: application developers and data managers (both as producers and users of SRs); information technology professionals and librarians; knowledge engineers and linguists (mostly producers); domain experts and researchers (producers, advanced users). A large number of tasks (e.g., creation, use, publication of a SR) and a great variety of tools were mentioned in the use cases. Typically, projects combine open source applications and ad-hoc in-house tools, while commercial solutions tend to support tasks for which no equivalent free open source software is available. Almost half of the use cases mentioned Semantic Web technologies, especially the use of triple stores (i.e., databases specialized in the storing and querying of RDF triples).

We analyzed the tasks mentioned across the different use cases, particularly regarding requirements that should be supported or enabled by tools. Four main areas and corresponding needs emerged: (i) Tools supporting **creation and maintenance of SRs** should include different types of users performing different editing phases, favoring visual and collaborative approaches, and allowing connection to existing SR catalogs or repositories so that they can be easily integrated and re-used. (ii) **Mapping between SRs** (i.e., correspondences between two SRs) should be supported by tools with an intuitive user interface and implementing state-of-the-art mapping algorithms, together with best practices guidance in the mapping process and a minimum of validation for the mappings created. Ideally, mapping should be also supported for non-SRs (e.g., via spreadsheets). (iii) An effective **use of SRs in data related applications** requires lowering the barrier for non-semantic experts by, for instance, using human-readable labels rather than only identifiers, complemented with services and metrics on SRs usage. (iv) Finally, **discoverability and availability of SRs** aligned with the F and A in the FAIR principles which involves means to create global, unique and persistent identifiers together with some associated metadata on provenance, format, version and so on. Whenever possible, such type of metadata should be generated automatically rather than relying on manual processes carried out by data curators (for example, information about data format, mapped resources, or previous versions of a given dataset, could be extracted automatically).

From our analysis on the use cases, we identified three main messages:

1. **Tools designed to work with SRs should be accessible to non-ontologists.** This implies that more attention should be paid to graphical interfaces, support for validation, and for methodological guidance in each task.
2. **Online platforms are needed** to lift the burden of local (or ad-hoc) installations and maintenance from users or individuals.
3. **Tools supporting common tasks involving SRs (e.g., editing, format conversion, mapping, and documenting) should be integrated, or integratable, to form flexible and interoperable workflows** to minimize the breadth of skills required to work with SRs.

---

[4] Respondents were asked to provide: 1) a title for their use case, 2) a problem statement, 3) requirements for SRs use, 4) information about their preferred semantic toolkit, 5) limits and expectations with working with semantic resources, 6) information about the data involved (type, format, storage, size, etc.), 7) manpower related information.

[5] Use cases were provided by GFAR, Univ. of Tor Vergata, INRAE (previous INRA & IRSTEA), AgroParistech, Embrapa, Ikerbasque Center for Climate Change, Univ. of Montpellier, AgMIP, CREA, CAAS, Solidaridad Network, Poznan Supercomputing and Networking Center, German Federal Institute for Risk Assessment, Univ. of British Columbia, ISKO, FSU Jena.

## *2.3 39 Hints to Facilitate the Use of Semantics for Data on Agriculture and Nutrition*

From the survey conducted (Section 2.2), it emerged that a variety of user profiles are involved along the different steps of the life cycle for semantic resources (Caracciolo et al., 2019b). We identified a set of recommendations distributed across five user profiles as summarized in **Table 1**. For the interested reader, we offer an extended version as part of the supplementary material.

## 3. Examples of Outputs Adoption

At the time of writing, the Agrisemantics WG is in "maintenance phase", with the goals of keeping the recommendations up-to-date, and promoting their adoption and refinement via public discussion on the recommendations through GitHub,[6] and a dedicated web site (Whitehead, 2020). Projects, institutions or initiatives qualify as adopters if they are actively implementing one or more of the recommendations, or they plan on doing so in the future. Adopters endorse the general view embodied in the recommendations, independently of any implementation ongoing or planned.

**Table 1:** Summary view of the recommendations (third column), organized by the user profile they are most relevant to (first column), and the topic they are concerned with (second column). In the third column, a reference to the corresponding recommendations if given in parenthesis.

| User profile | Topic | Recommendations |
| --- | --- | --- |
| Semantic web developers | Integration | Consider integration and interoperability from the beginning to facilitate connections to other SRs (R10) and integration into suites (R1). Using open source licenses (R2), adding metadata (R6), and automatically notifying of new releases (R11) will make integration easier |
| | Best practices and quality | Following best practices (R3) and ensuring some quality level (R9) are always a good idea when developing software |
| | Alignment/matching/mapping support | With that many vocabularies, it is useful to support alignment, matching, and/or mapping approaches (R12, R13). Also allow for customization of those editing and alignment tools supported by your platform (R8) |
| | End User | Always consider end-users, prepare to provide documentation (R4) and support for both semantic and non-semantic experts (R5). Also, include some guidance on how to choose the right semantic data structure (R7) |
| Semantic professionals | FAIR principles | Implement the FAIR principles (R14), deposit versions in repositories (R15) from alpha to production releases (R16), reuse SRs (R18, R20) |
| | Best practices and quality | Use best practices (R17) and metrics to assess usage (R22) |
| | Community | Promote communities of practice (R19) and use of standards (R21) together with recommendation and training (R23) and use cases, lessons learnt, etc. (R24) |
| Developers of data platforms consuming semantic resources | Technologies | Keep up to date concerning technologies (R25), support multiple formats (R29) |
| | Best practices and quality | Support multiple functionalities (R26), use metadata to describe data (R27) and semantics to characterize it (R28) |
| Data managers and producers | Best practices and quality | Get familiar with Research Data Management (RDM) (R30) plans and use them (R31), carefully characterize your data (R32), prefer SRs whenever possible (R33), and document cross-references (R34) |
| Policy makers and funders | Best practices and quality | Encourage the use of SRs (R35), |
| | Maintenance | Provide support for maintenance (R36, R38) |
| | Community | Promote discoverability (R37) and dissemination via training (R39) |

---

[6] https://github.com/agrisemantics/recommendations/issues.

To date, 14 institutions or initiatives have recognized themselves as adopters of the Agrisemantics output.[7] Here, we give an overview of the adopters so far, describing four adoption cases in more detail. These four cases are representative of the various ways in which the recommendations can be implemented – a large portal for SRs (AgroPortal), an initiative aimed at lifting domain specific SRs with Semantic Web technologies (Caliper), an RDF and OWL files editor (VocBench), and a platform enabling reasoning and data integration based on semantic technologies (k.LAB and ARIES).

The Scientific and Technical Information Department (DIPSO) of the French National Research Institute for Agriculture, Food and the Environment (INRAE) is heavily committed to developing semantic resources like ontologies and thesauri for data related uses within the organization. In particular, INRAE's laboratory of Clermont-Ferrand, Technologies and information Systems for agricultural systems (TSCF), works on the use of sensor data – represented as linked open data– to improve agricultural practices and decision support (Nguyen et al., 2020; Roussey et al., 2019). The two National Research Agency supported projects, D2KAB and FooSIN,[8] respectively headed by University of Montpellier and INRAE, are explicitly relying on Semantic Web technologies and SRs in their processes from transforming agrifood data to knowledge. Two Dutch centers, the Wageningen Data Competence Centre (WUR) and the Dutch Techcentre for Life Science (DTL), are early supporters of data sharing and interoperability in agriculture, and implementers of recommendations on tools interoperability in the area of infrastructures for open science. The Global Open Data for Agriculture and Nutrition (GODAN), an international initiative aimed at promoting the production and sharing of open data in agriculture and nutrition, supports Agrisemantics[9] The Italian Council for Agricultural Research and Economics (CREA) is interested in the application of the Semantic Web technology stack (starting with RDF, including standard modelling, use of global identifiers) to improve the production and sharing of soil maps. Finally, the GO FAIR Food Systems Implementation Networks is planning on adopting the Agrisemantics.[10]

### 3.1. AgroPortal

AgroPortal is a public repository of semantic resources (*AgroPortal, release 2.0*, Sept. 2020; Jonquet et al., 2018b, 2018a) reusing the NCBO BioPortal technology (Noy et al., 2009), and offering ontology hosting, search, versioning, visualization, comment and recommendation. It enables semantic annotation and supports basic alignment features via a user web interface and web services. At the time of writing, the platform hosts 128 SRs, ranging from reference domain ontologies to important thesauri used in agrifood. AgroPortal primarily addresses the first three groups of identified user profiles, however, the platform is easily accessible for data managers too. We overview hereafter AgroPortal implementation of the Agrisemantics recommendations, for more details see Annex II.

AgroPortal relies on semantic web technologies and aims to facilitate discoverability and long-term availability of SRs (R37, R25 and R38); it uses RDF as a pivot language and transforms non-RDF SRs to alternative RDF files available for download and stored them in an RDF triple store (R29). AgroPortal is endowed with a REST web service API serving XML or JSON-LD content and facilitating automatic use of SRs and related services in data workflows (R1). Based on those services, content search, indexing and annotation tasks are supported (R27). Specific agrifood communities (e.g., INRAE, Crop Ontology project, RDA WDI WG) are supported with customized facilities, such as SR grouping and interfaces (R19).[11] AgroPortal supports SR developers in using a unified metadata model based on a shared specification (MOD 1.4 (Jonquet et al., 2018b)) (R3) and, when available, metadata can also be automatically generated (R6). AgroPortal's SR metadata are mainly used to: (i) serve as a means to create connections to other existing catalogs, libraries and repositories (FAIRsharing, OBO Foundry, Ontobee, etc.) (R10); (ii) assess SRs via analytical metrics (user reviews, projects, usage description, etc.) (R22); (iii) help in determining the level of FAIRness of SRs (IDs, language, license, provenance, etc.) which in the future will be done automatically. As a result, it will allow SR developers to make their SRs more FAIR and SR users to identify FAIR SRs (R14). Finally, AgroPortal supports basic ontology alignment functionalities and in the future it will support collaborative mapping evaluation (R12). It pursues the vision of becoming a reference open platform for SR in agrifood by synchronizing efforts with other SR repositories, especially the ones relying on the NCBO BioPortal technology (R35).[12]

---

[7] The up-to-date list of adopters can be found at: https://agrisemantics.org/#adopters/.
[8] www.d2kab.org and https://ist.blogs.inrae.fr/foosin/.
[9] https://www.godan.info/pages/godan-working-groups.
[10] https://www.go-fair.org/implementation-networks/overview/food-systems/.
[11] To access only these SRs one can go to the specific slice: http://inrae.agroportal.lirmm.fr/ontologies.

### 3.2. Caliper

Caliper is a platform for sharing machine-friendly versions of agrifood-related statistical classifications ("Caliper" n.d.). Being mostly an effort of conversion of existing statistical classifications into linked open vocabularies, the recommendations most relevant to Caliper are those addressing semantic professionals. Moreover, Caliper provides inputs to developers of tools consuming SR and data managers, in that the classifications made available through the platform are oriented to use in information systems.

The semantic professionals working on Caliper have reused common metadata schemes, endowed with adequate documentation (R14), and adopted existing best practices for modelling and using SRs (R17), as well as for the development of new ones. Also, existing resources have been widely reused (R18), communities of practice have been promoted (R19), and domain-specific standards for alignments consistently reused (R21). Caliper is committed with the communication of the benefits of semantics (R24). All tools used are open source (R2), supporting best practices regarding the modelling and access of a SR (R3), aimed at supporting non-semantic experts working with SRs (R5) and with a certain degree of customization allowed (R8).

Ongoing work is devoted to adapting some of the recommendations to the specific applications covered by Caliper, for example the definition of appropriate licenses schemes (R23), also in collaboration with its community of interest (R19). Collaboration is ongoing between semantic professionals and tool developers re customization (R8) and connection to domain repositories (R10), implementing mechanisms for new versions. Mappings between resources are widely present, either based on correspondence tables defined and approved by classifications custodians, or enriched by semi-automatic mechanisms.

### 3.3 VocBench 3

VocBench 3 ("VocBench" n.d.) is a free and open source (R2) advanced collaboration environment for creating and maintaining ontologies, thesauri, code lists, authority tables, lexicons and link sets in compliance with Semantic Web standards recommended by the W3C (R25, 26, 27 and 28) (Stellato et al., 2015). VocBench is used to maintain vocabularies and ontologies in a wide number of domains,[13] including the agrifood sector (e.g., at FAO, it is used to maintain AGROVOC and the vocabularies in Caliper; at INRAE, it is used to maintain vocabularies on ecosystems and biodiversity). The VocBench 3 project is funded by Action 1.1 of the ISA2 Programme of the European Commission for "Interoperability solutions for public administrations, businesses and citizens". VocBench site contains documentation, download links and other references (R4).

VocBench fully meets the established requirements for data editing and consumption tools, supporting the profiles defined in Section 2.3 (semantic professionals, data managers, policy makers) in fulfilling their job by respecting the requirements associated with their figures. As a software suite, many of the tools offered by VocBench are available as separate components that can be integrated in other software (R1). Depending on the standard being adopted and the type of user, the system provides several user-tailored facilities for easily modeling the needed resource (R5, R7, R26) and toggles or makes optional (with properly conceived default settings) various features (R8). Quality checking (R9) is provided by dedicated Integrated Constraint Validation tools (ICVs) and by support for SHACL shapes.

### 3.4 k.LAB and ARIES

The software stack k.LAB (Villa et al., 2017, 2014) aims at providing an infrastructure to integrate distributed scientific knowledge and data. It relies on the "semantically integrated modelling", a scientific practice that reconciles rigorous formal semantics with the production, use and curation of scientific artifacts such as datasets, mathematical models and computational services. The key feature of Integrated Modelling is that it distinguishes the process of "observation" from the resulting "data" – consequently, the logical representation of the "observation" remains distinct from both the actual data produced, and the functional knowledge usually implemented in mathematical models and algorithms. ARIES (ARtificial Intelligence for Ecosystem Services, (Villa et al., 2014)) is the flagship project based on the k.LAB technology, addressing decision-makers and researchers. As such, k.LAB and ARIES implement recommendations for all user profiles considered.

---

[12] For information on the technology see the OntoPortal Alliance (https://ontoportal.org).
[13] For a list of current user of VocBench, see http://vocbench.uniroma2.it/support/community.jsf.

The SRs editing environment included in k.LAB allows for export in OWL2, a standard W3C language (R1), it is licensed as an open source software (R2) and supports best practices to realize the FAIR principles (R3). Thorough documentation for the platform will be made available in the next release (R4), while metadata and documentation for the SRs are automatically generated both in formal and natural language (R6). k.LAB includes a web-based portal, k.Explorer, geared towards non-technical users (R5). In developing SRs, users are guided by the Integrated Modelling methodology, also supported by automatic check for logical constraints and quality (R7, R8, R9). k.LAB allows for connection to third-party resources ("authorities") (R10) – in particular, the ARIES project uses the platform Caliper as a source of standard, widely accepted vocabularies covering agricultural concepts, plus other "authorities" such as IUPAC (chemical species) and GBIF (taxonomy). All content is network-available and automatically synchronized to users (R11). Alignment features supporting OWL2 formats are under development (R12), although SPARQL facilities are not yet in place (R13). All ontologies used and developed in k.LAB are available in version-control enabled repositories (R15) and different branches can be made available to different user groups (R16). The FAIR principles are adopted as its overarching goal, and several well-established resources are consistently reused (R17, R18). An Integrated Modelling Partnership backs up k.LAB (R19) and a summer school is regularly held (R23). k.LAB also implements algorithms for ranking SRs based on usage data (R22) and it offers functionalities aimed at discovering and reusing datasets (R26, R27, R28), according to the SRs used in them (R32), and for typing data (R33).

## 4. Discussion and Conclusions

Semantics is part of any data-related task. All data scientists know the importance of good and unambiguous definitions of data dimensions, crucial to all phases of data analysis. However, semantics is often left implicit in the data, the semantic resources used to create the data are not easily accessible, or available in non-standard formats, non(easily) machine-readable – all factors hampering the possibility of reusing data in information systems or integrating it with other datasets and ultimately limiting the interoperability of data. The work reported in this paper aims at contributing to agricultural data interoperability, by providing operational suggestions to the user profiles involved in the entire data lifecycle, with a special focus on creators and users of semantic resources.

While convinced of the intrinsic domain-independence of semantics as a way to express the meaning of data and utilize them in computer-based applications, we adopted a community-driven approach to identify the needs of the community, distill them into high-level recommendations, and promote action within the community. Therefore, much emphasis is put on tools, for example insisting on their capability of accessing domain specific repositories of data and SRs. Also, emphasis is put on making tools usable by different user profiles, including domain experts, and accessible also to institutions where funding for IT development is limited (by encouraging the development of web-based tools).

The view embodied in these recommendations has many points of contact with the FAIR principles (GO FAIR, n.d.; Wilkinson et al., 2016), as shown in **Table 2**. For example, recommendation (R14) explicitly urges semantic professionals to make their SRs FAIR, by providing persistent identifiers to resources, sharing them in public repositories and catalogues, and reusing common metadata schemes. However, some aspects of our recommendations extend the FAIR data principles: e.g., develop awareness and training for semantics.

One of the points of convergence between our recommendations and the FAIR principles is the role of community agreements for achieving an acceptable level of "FAIRness" (e.g., FAIR Principle R1.3), for example for what concerns the adoption of specific metadata schemes. Community agreement is important to minimize the effort associated to data reuse, and to avoid "isolation" of resources – e.g. in those cases when different level of richness of metadata reflect different technical or organizational capacity of the data curator (see (David et al., 2020; Schultes, 2019)). However, the recommendations elaborated from the Agrisemantics WG also cover aspects not explicitly considered within the FAIR data principles, such as the importance of promotion and training (R19, R23, R24, R39)) or the recommendations addressed to policy makers (R35, R36, R38). In our set of recommendations, we have also drawn an emphasis on tools intelligibility and desired functionalities to address multiple aspects of FAIR in the SR lifecycle (especially R1-R14). Also, our recommendations stress the needs of different user profiles not only data stewards which are the primary targets of the FAIR principles.

**Table 2:** How the Agrisemantics WG's recommendations align with the FAIR principles.

| FAIR | Agrisemantics | Comment |
| --- | --- | --- |
| Findable | R6, R10, R14, R16 | Our recommendations insist on the importance of rich standardized metadata for SR (F2) and to register or index SR in searchable catalogs and repositories (F4). Also, we do recommend SR are assigned a globally unique and persistent identifier (F1). |
| Accessible | R10, R14 | Several of our recommendations enforce the use of Semantic Web technologies to facilitate data access (A1). We also encourage sharing both SR and their metadata in various repositories and catalogs so they become accessible, even when the SRs are no longer directly available (A2). |
| Interoperable | R1, R7, R12, R13, R18, R21, R26, R28, R29, R31 | A handful of our recommendations are related to interoperability. Recommendations related to format and technologies encourage the use of appropriate, formal, accessible, shared, and broadly applicable languages for knowledge representation, alignment, and metadata (I1). We also strongly support the (re)use of other SRs and/or standard vocabularies when building a SR (I2). Then, the recommendations about alignment intersect with the notion of qualified references between SR (I3). |
| Reusable | R2, R4, R26, R32, R33, R36 | We do recommend tools and SRs are released with a clear and accessible data usage license (R1.1) and work with domain-relevant community standards for both SR content and metadata (R1.3) as well as with semantically enabled data types. We also encourage important documentation and description of the SRs and provenance processes (R1.2). |

The adoption stories we reported in this paper (Section 3) give a sense of the wide range of applications of the recommendations of the Agrisemantics WG. AgroPortal represents the importance of domain-specific repositories and tools for mappings; Caliper addresses the issue of levering existing semantic resources to an adequate level of formalization and standardization to improve interoperability of statistical data. VocBench addresses the needs of semantic professionals, offering a web-based platform for the creation and maintenance of semantic resources according to best practices. k.LAB enables effective data integration with a sound and full-fledged semantically enabled platform. Future activities of the group include moving further and deeper in the two directions of detailing further the recommendations and promoting larger adoption.

# Additional Files
The additional files for this article can be found as follows:

- **Annex I.** Hints to Facilitate the Use of Semantics with Agricultural Data. DOI: https://doi.org/10.5334/dsj-2020-047.s1
- **Annex II.** Mapping Adoption Cases and Recommendations. DOI: https://doi.org/10.5334/dsj-2020-047.s2

# Acknowledgements
The Agrisemantics Working Group thanks RDA and RDA Europe for their support.

Caterina Caracciolo was supported by the Bill and Melinda Gates Foundation, the Food and Agriculture Organization of the UN and the RDA Europe Ambassador Programme. The views expressed in this publication are those of the author and do not necessarily reflect the views or policies of the Food and Agriculture Organization of the United Nations, nor those of the other funding bodies.

Brandon Whitehead acknowledges with thanks the support of the CABI Development Fund. CABI is an international intergovernmental organization and we gratefully acknowledge the core financial support from our member countries (and lead agencies) including the United Kingdom (Department for International Development), China (Chinese Ministry of Agriculture), Australia (Australian Center for International Agricultural Research), Canada (Agriculture and Agri-Food Canada), Netherlands (Directorate-General for International Cooperation), and Switzerland (Swiss Agency for Development and Cooperation). See https://www.cabi.org/about-cabi/who-we-work-with/key-donors/ for details.

Sophie Aubin, Clement Jonquet, Emna Amdouni, Romain David and Catherine Roussey were supported, in part, by the French National Research Agency (ANR) Data to Knowledge in Agronomy and Biodiversity (D2KAB – www.d2kab.org – ANR-18-CE23-0017). Romain David was partly supported by the EPPN2020 project (H2020 grant N°731013), the EOSC-Life european program (grant agreement N°824087), the 'Infrastructure Biologie Sante' PHENOME-EMPHASIS project funded by the French National Research Agency (ANR-11-INBS-0012) and the 'Programme d'Investissements d'Avenir'.

This manuscript publication is supported by the RDA Europe 4.0 project that has received funding from the European Union's Horizon 2020 research and innovation programme under grant agreement No 777388.

## Competing Interests
The authors have no competing interests to declare.